\documentclass[journal]{IEEEtran}

\usepackage{cite}

 \usepackage{multirow}
 \usepackage{colortbl}
 \usepackage{hhline}

\usepackage{caption}
\usepackage{subcaption}

\usepackage{algorithm}
\usepackage{algorithmic}

\usepackage{color,soul}
\usepackage{gensymb}
\usepackage{dsfont}

\ifCLASSINFOpdf
  \usepackage[pdftex]{graphicx}
  \graphicspath{{../pdf/}{../jpeg/}}
  \DeclareGraphicsExtensions{.pdf,.jpeg,.png}
\else
  \usepackage[dvips]{graphicx}
  \graphicspath{{../eps/}}
  \DeclareGraphicsExtensions{.eps}
\fi

\usepackage[cmex10]{amsmath}
\usepackage{amssymb}

\usepackage{euscript}

\usepackage{multirow}
\usepackage{algorithm}
\usepackage{diagbox}
\usepackage{physics}
\usepackage{tikz}

\usepackage{amsthm}
\usepackage{bm}

\usepackage{url}

\begin{document}
%
\title{Data-Driven Outage Restoration Time Prediction via Transfer Learning with Cluster Ensembles}
%
%

\author{Dingwei~Wang,~\IEEEmembership{Graduate Student Member,~IEEE,}
	Yuxuan~Yuan,~\IEEEmembership{Graduate Student Member,~IEEE,}
	Rui~Cheng,~\IEEEmembership{Graduate Student Member,~IEEE,}
	and Zhaoyu~Wang,~\IEEEmembership{Senior Member,~IEEE}
\thanks{This work was supported in part by the National Science Foundation under EPCN 2042314 and CMMI 1745451, and in part by Advanced Grid Modeling Program at the U.S. Department of Energy Office of Electricity under Grant DE-OE0000875. (\textit{Corresponding author: Zhaoyu Wang})

D. Wang, Y. Yuan, R. Cheng, and Z. Wang are with the Department of
Electrical and Computer Engineering, Iowa State University, Ames,
IA 50011 USA (e-mail: dingwei@iastate.edu; wzy@iastate.edu).
 }
}

%
%

\markboth{Submitted to IEEE for possible publication. Copyright may be transferred without notice}%
{Shell \MakeLowercase{\textit{et al.}}: Bare Demo of IEEEtran.cls for Journals}

%



\maketitle

\begin{abstract}

This paper develops a data-driven approach to accurately predict the restoration time of outages under different scales and factors. To achieve the goal, the proposed method consists of three stages. First, given the unprecedented amount of data collected by utilities, a sparse dictionary-based ensemble spectral clustering (SDESC) method is proposed to decompose historical outage datasets, which enjoys good computational efficiency and scalability. Specifically, each outage sample is represented by a linear combination of a small number of selected dictionary samples using a density-based method. Then, the dictionary-based representation is utilized to perform the spectral analysis to group the data samples with similar features into the same subsets. In the second stage, a knowledge-transfer-added restoration time prediction model is trained for each subset by combining weather information and outage-related features. The transfer learning technology is introduced with the aim of dealing with the underestimation problem caused by data imbalance in different subsets, thus improving the model performance. Furthermore, to connect unseen outages with the learned outage subsets, a t-distributed stochastic neighbor embedding-based strategy is applied. The proposed method fully builds on and is also tested on a large real-world outage dataset from a utility provider with a time span of six consecutive years. The numerical results validate that our method has high prediction accuracy while showing good stability against real-world data limitations.

\end{abstract}



\section{Introduction}\label{introduction}

Recent Texas blackout has demonstrated the danger and inconveniences people face during a severe power outage event. In general, power outages have significant impacts on production, transportation, communication, and health supply service, resulting in significant economic losses. When an outage occurs, utilities need to make a series of decisions quickly, including detecting and locating the fault, estimating costs and the number of customers affected, predicting outage restoration time, planning repair strategies and dispatching crews \cite{jaech2018real}. From the customer's perspective, the most important and concerned information is timely and accurate outage recovery time prediction, which will greatly help them plan for subsequent arrangements in advance. However, it is challenging to estimate outage recovery time as power outages are typically unplanned with limited information. Moreover, the causes of power outages involve a wide variety of factors, e.g., bad weather, human behaviors, and equipment failures \cite{modelingweatherrelated, outageevents,adaboost,outagecause}. To improve customer satisfaction, accurate prediction of outage recovery time is becoming a top priority for utilities.

In recent years, research studies have focused on interring the quantity and duration of outages by using various approaches and data sources, which can be broadly classified into two groups based on prediction targets: \textit{Group I -} Prediction of outage \textit{duration}. The authors in \cite{timeofoutage} analyzed restoration duration of outages using statistical and quantitative methods with several factors including time of the outage, consequences, and environmental conditions. In \cite{liu2007statistical}, an accelerated failure time model using severe weather records were developed to estimate duration of outages. In \cite{predictiondeeplearning}, the authors summarized six years' historical outage data and proposed a deep neural network to predict repair and restoration time with respect to severe weather events. In \cite{yue2017bayesian}, the authors utilized radar observation data and further proposed a generalized weather condition dependent failure rate model, based on the Bayesian prediction algorithm, to provide an prediction of outage numbers. \textit{Group II -} Prediction of outage \textit{numbers}. In \cite{domijan2005effects}, the authors performed a Poisson regression model to predict average number of outages over a period under normal weather conditions. The authors in \cite{kankanala2012estimation} estimated distribution system outages numbers caused by wind and lighting using an artificial neural network. In \cite{owerko2018predicting}, a graph neural network is proposed to predict power outages numbers by utilizing weather variables. A summary of the literature is shown in Table \ref{tab:review}.

\begin{table*}[ht]
\caption{LITERATURE REVIEW ON OUTAGE PREDICTION IN DISTRIBUTION SYSTEMS}
\centering
 \begin{tabular}{c}
 \includegraphics[width=0.97\textwidth]{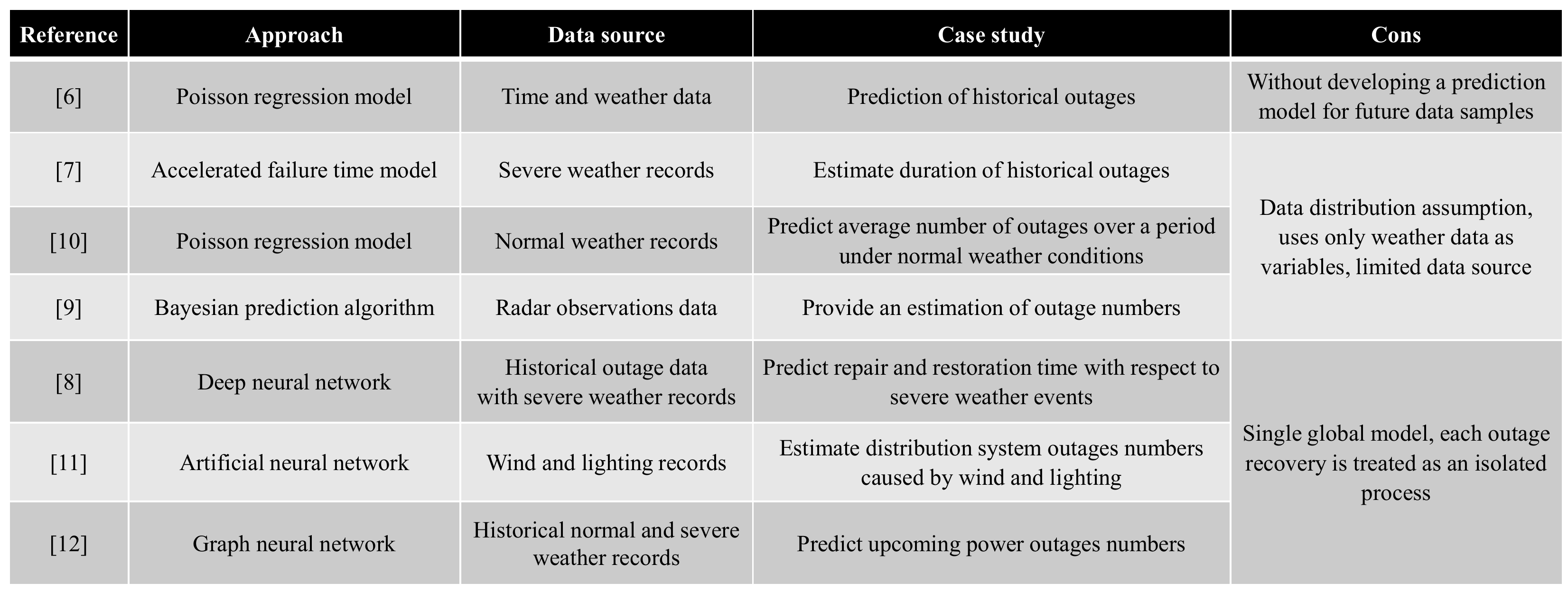} 
  \end{tabular}
  \label{tab:review}
\end{table*}

Even though the previous works demonstrate valuable results, some research challenges remain outstanding in this area. First, most studies in group I are generally based on the assumption that each outage recovery can be treated as an isolated process. In other words, with respect to multiple outages that occur in the same systems, the existing methods estimate the restoration time separately without consideration of the correlation among multiple coinciding outages. However, in actual grids, while maintenance crews are repairing an outage, the repair of another outage occurring in the neighboring area may be delayed due to crew shortages. Thus, such an assumption in Group I reduces the accuracy of these prediction models. Second, some studies in group II trained a global model for the whole historical outage dataset, which ignores the uncertainty caused by the heterogeneity of outage events under different scales and factors, thus reducing prediction performance for unseen outages. Also, given the paucity of real outage reports, most existing works rely on the weather information to develop their models and assume the weather conditions across each small area are homogeneous. Nevertheless, while severe weather is one of the leading causes of power outages, other factors, such as equipment failure and human factors, should not be ignored.

\begin{figure}[tbp]
 	\centering
 	\includegraphics[width=3.4in]{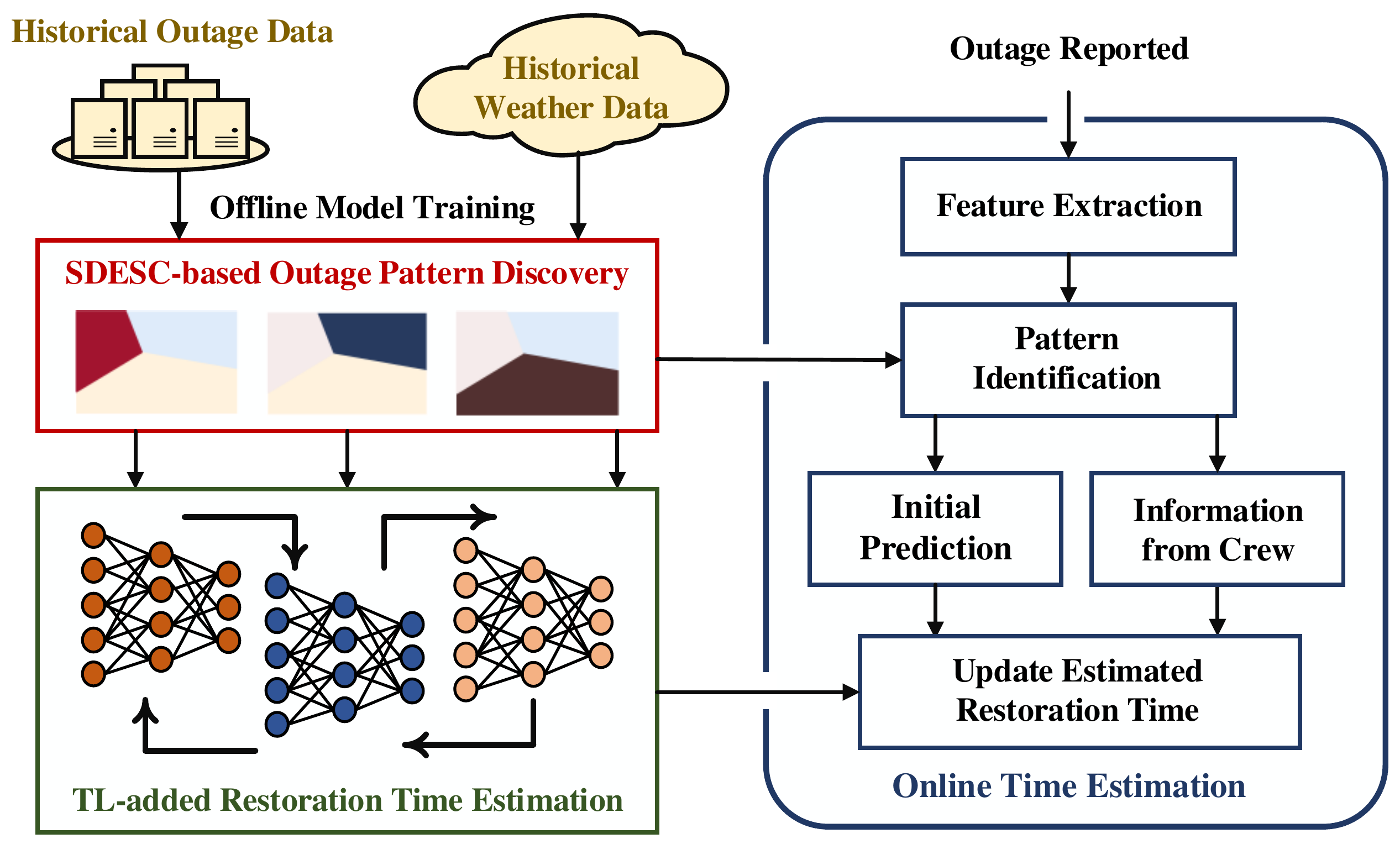}
 	\caption{The flowchart of the proposed method.}
 	\label{fig:flow}
\end{figure}

To this end, we propose a novel data-driven method to predict outage restoration time using transfer learning with cluster ensembles. The flowchart of our proposed method is depicted in Fig.\ref{fig:flow}. Compared with the existing studies, the contributions of our work focus on: 1) To investigate the interaction of simultaneous outage events during a period, we extract the statistics from real-world outage reports and calculate the cumulative number of coinciding outages and affected customers. The temporal information in outage data, such as the start and end time and restoration time, is utilized to summarize the real-time numbers of outages and customers affected in a time span. This information can be explored to approximate the system stress for prediction model developments. 2) Unlike previous methods that train a global prediction model, the proposed method estimates the restoration time in a cluster-wise manner to deal with the uncertainty caused by the heterogeneity of outage events. Specifically, a sparse dictionary-based ensemble spectral clustering (SDESC) method is developed to efficiently group the historical outage events. For each data subset, a prediction model is trained to construct an end-to-end mapping between the outage-related information and the restoration time by leveraging deep learning techniques. 3) Based on our investigation of real-world outage datasets, there may be huge gaps between the amount of data available for different patterns and scales of outages. Therefore, a transfer learning strategy is integrated with the proposed prediction framework to avoid the overfitting risk caused by the data scarcity of the specific outage patterns. 4) The proposed approach leverages not only high-level weather information but also exploits outage-related features collected by our utility partner, which are time and location of outages, number of customers interrupted, cause code, and duration of repair/restoration process. The proposed prediction methodology has been tested and verified using real outage data.

The rest of this paper is organized as follows: Section II introduces the problem definition and describes the available outage dataset in detail. Section III proposes the sparse dictionary-based ensemble spectral clustering method. In Section IV, the transfer learning-added outage restoration time prediction model is presented. The numerical results are analyzed in Section V. Section VI concludes the paper.

\section{Problem Definition $\&$ Outage Data Description} \label{descreption}

\subsection{Problem Definition}
When an outage occurs, one of the most common questions that customers may ask is: How
long will it take to restore the power\cite{timeofoutage}? If utilities can answer this question at an early stage, customers can plan ahead and accordingly to avoid inconvenience caused by power outages. In practice, outage restoration time is defined as the time from the start of the outage to the service is fully recovered to the customers \cite{predictiondeeplearning}. In actual grids, one common solution is to formulate a mathematical relationship between the restoration time and the number of interrupted customers \cite{timeofoutage,liu2007statistical,yue2017bayesian}. Such a solution embodies a set of assumptions, such as the pre-define statistical models with fixed parameters (e.g., Poisson distribution). However, the distribution of outage restoration time may not follow the pre-defined pattern; meanwhile, some variables and features cannot be considered in the statistical models due to the curse of dimensionality, which reduces the accuracy of the prediction. Hence, machine learning-based methods are receiving increasing attention due to the unprecedented amount of data collected by utilities. Basically, the learning-based solution is based solely on using historical outage data without an assumption of data distributions to develop a supervised prediction model. 

\subsection{Available Outage Dataset}
\begin{figure}[tbp]
 	\centering
 	\includegraphics[width=3.4in]{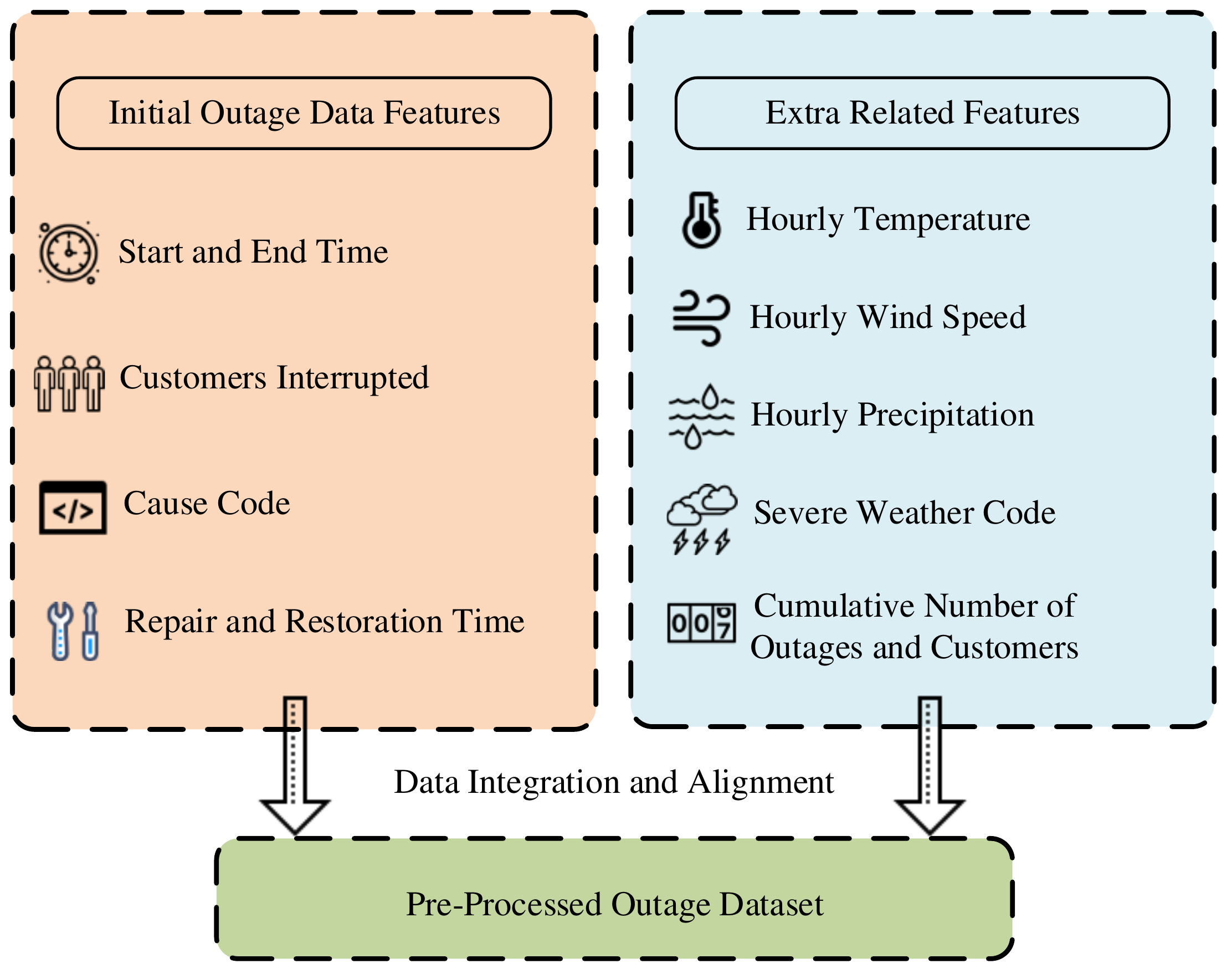}
 	\caption{The description of the outage dataset.}
 	\label{fig:dataset}
\end{figure}

The outage data under study includes over 16,000 records over a six-year period in New York State. Fig. \ref{fig:dataset} describes the structure of the available outage dataset. The original information on each outage record includes: start and end time of the outage accurate to seconds, number of customers interrupted, repair and restoration time accurate to seconds, cause code, location, and distribution network circuit number. Specifically, the start and end time of the outage events are reported by fuse cards that record the time at which an outage starts and ends according to the loss of power. The cause information is summarized into two types: \textit{Cause key} - There are 63 causes, of which 5\% of the records are weather-related, another 14\% are animal-related, and the rest are mainly due to component malfunctions, tree limbs, and debris. The top two causes of all outage events are tree limbs near the clearance zone of equipment and squirrels. The top weather-related causes are precipitation, wind, and lightning strike. \textit{Equipment cause key} - The outages are caused by equipment failure. The top equipment-related causes are system failure, conductor disconnection, and transformer malfunction. These two types of cause information are recorded in each outage and represented by a digital code. 

In this work, data preprocessing includes two steps: 1) \textit{Missing and bad data cleaning} - Given equipment failures or human mistakes, missing and bad data in historical outage datasets are typically unavoidable and can lead to misclassification, which negatively affects the performance of data-driven methods. Hence, the available dataset is initially processed to clean missing and bad outage samples. Data samples with empty entries are removed first. Then, following the engineering intuition, data samples with logically incorrect entries (e.g., the restoration time is greater than total outage time) and grossly erroneous values (e.g., the restoration time is extremely illogically high) are removed. 2) \textit{Outage-related features investigation} - Leveraging cross-domain insights from public weather data and the geographic information of systems, the raw dataset is explored by adding hourly temperature, precipitation, and wind speed. These data are collected from the National Oceanic and Atmospheric Administration (NOAA) \cite{noaasevere, noaa}. The hourly data from the NOAA is aligned with each outage sample based on the start time of each outage. Given that severe weather events play a crucial role in outage restoration time prediction, the weather condition of each outage record is also marked as a discrete code, consisting of normal conditions, snowstorm, lightning strike, high-speed wind, and flood \cite{jaech2018real}. Other features we have added are \textit{cumulative number of coinciding outages and number of customers affected}. Specifically, in this work, the cumulative number of coinciding outages is the number of outages presented at a certain period that has not yet been resolved. This number varies over time while the outages occur and are restored. When this number remains near zero, it indicates that the system is in normal condition prior to outages, or that outages happened infrequently and are resolved quickly. Conversely, under certain stressed conditions, this number can be relatively high, indicating that numbers of outages have stacked and affected the restoration time. For example, on December 10th, 2014, the cumulative number of coinciding outage events was over 100, and the average outage restoration time was above 300 minutes. In contrast, the cumulative number of coinciding outage events is relatively low on April 15th, 2015, the average outage restoration time is below 60 minutes.

\begin{figure}[tbp] 
	\centering
	\includegraphics[width=3.4in]{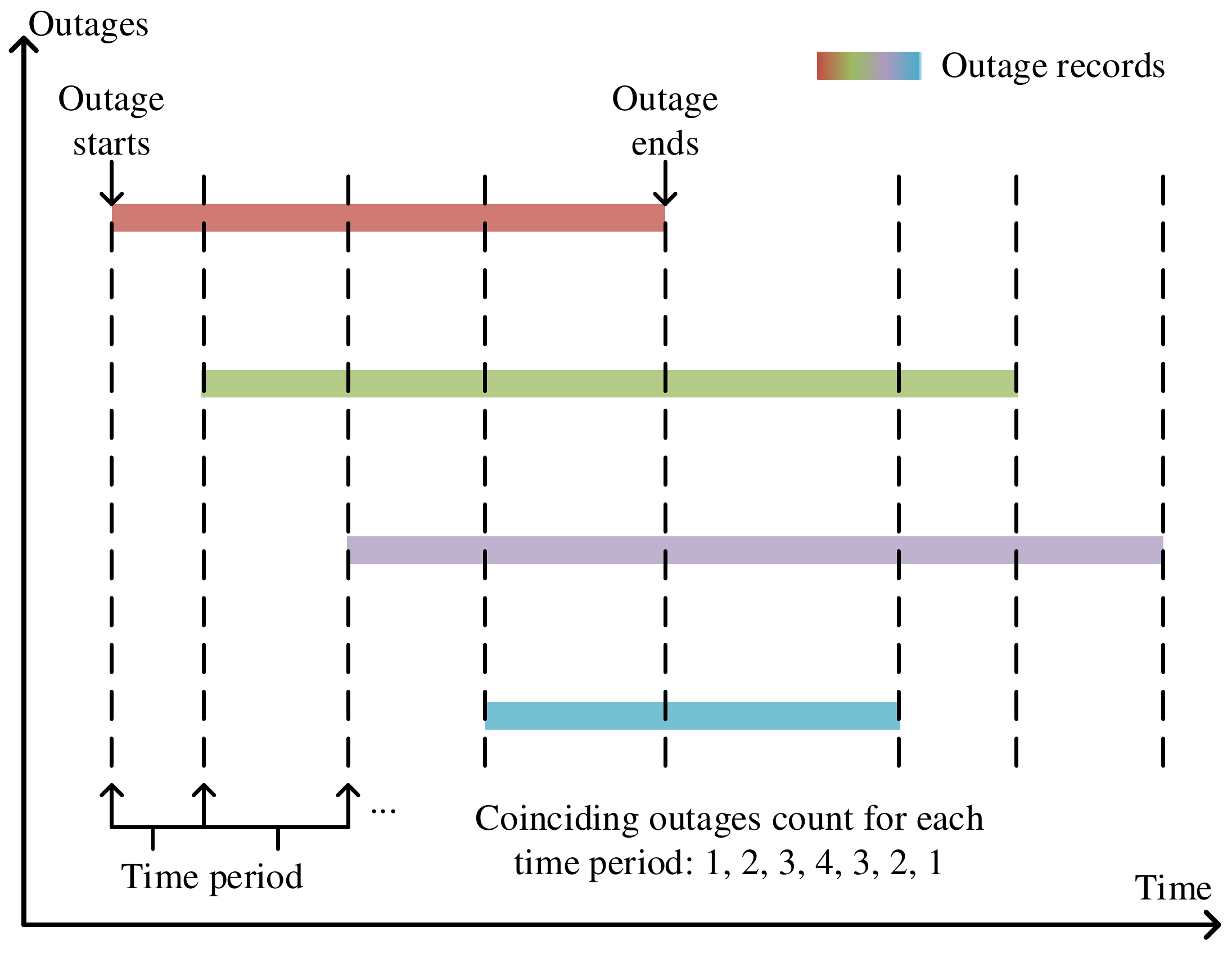}
	\caption{Exemplary of the cumulative number of coinciding outages.}
	\label{fig:cummulative}
\end{figure}

Fig. \ref{fig:cummulative} demonstrates an example for calculating the cumulative number of coinciding outages. Each dashed line on the graph is a timestamp that is recorded at each start and end of each outage. By comparing each two adjacent dashed lines, a specific time period can be defined to explore the cumulative number of coinciding outages by counting the numbers of outages. Following this process, the cumulative number of coinciding outages for each period is calculated by sorting the timestamps by their orders of occurrence:
\begin{equation}
\label{eq:cumulative}
\begin{aligned}
    C_{outages} = {C_{o}^{t_i}} - {C_{r}^{t_i}}
\end{aligned}
\end{equation}
where $C_{o}^{t_i}$ is the cumulative total outages at time $t_i$, $C_{r}^{t_i}$ is the cumulative total restorations at time $t_i$, and $C_{outages}$ is the number of simultaneous outages at time $t_i$. The cumulative number of coinciding outages is used to define the stress of the system and can provide additional dimension information for outage grouping, as well as enhance the variability of the metric. Similar to the above definition and \eqref{eq:cumulative}, the cumulative number of customers interrupted can be calculated by replacing the number of coinciding outages with the number of customers.

\section{Historical Outage Data Discovery using SDESC} \label{data}

Nowadays, utilities are constantly attempting to collect as much information as possible on power outages. However, the vast majority of outages in distribution systems are small-scale and medium-scale; in contrast, large-scale outages are still rare, thus leading to a data imbalance problem\footnote{The data imbalance problem refers to an unequal distribution of classes in the training dataset. When the dataset is imbalanced, the trained model typically fails to capture the hidden features of minority groups. Thus, the performance of a supervised model may suffer from the fact that the distribution of the target variable is skewed.} \cite{faultcausewithimbalanced}. In this case, it could result in overfitting problems when the raw outage data is used to train a global prediction model.

To address the challenge posed by the real-world imbalanced outage dataset, a novel unsupervised method, known as sparse dictionary-based ensemble spectral clustering (SDESC), is leveraged to distinguish the hidden outage features and partition the historical dataset into distinct subsets. The proposed method follows the line of unsupervised research utilizing the spectral analysis to discover the 
hideen features. Furthermore, the sparse coding technique is adapted to decrease the complexity of outage event-based adjacency matrix construction and eigen-decomposition, thus greatly reducing the cost of practical implementation \cite{LSC}.

Regarding notations, let ${\bm O}=[{\bm O}_1,\bm{O}_2,...,\bm{O}_m]$ represents the historical outage dataset, where $m$ is the total number of the outage events, $\bm {O}_i\in\mathbb{R}^{n \times 1}$ represents the outage-related information in the $i$-th outage event including $n$ features. And let ${o}_{ij}$ denotes the $i$-th row and $j$-th column entry in $\bm {O}$. In actual grids, given the accumulation of historical outage data over many years, $m$ is a large real number. Sparse coding is utilized to find a sparse representation of $\bm O$ that consists of a dictionary ${\bf D} \in \mathbb{R}^{m \times p}$ and a representation ${\bf R} \in \mathbb{R}^{p \times n}$ \cite{LSC}. Specifically, each column of $\bf D$ contains the features of the outage samples. Each column of $\bf D$ represents the $p$-dimensional (i.e., $p \ll m$) representation with respect to the raw data samples in $\bf R$. To minimize the approximation error, this process is regulated as a optimization problem using the Frobenius norm:
\begin{subequations}\label{eq:sparse}
\begin{align}
\underset{{\bf D},{\bf R}}{{\rm minimize}}\hspace{2mm}&||{\bm O}^{T}-{\bf D}{\bf R}||_F^2\\
{\rm subject}\,{\rm to}\hspace{2mm}
&||\bf D_i||_F^{2}<= C, \forall i = 1,...,p.
\end{align}
\end{subequations}

Obviously, solving the above optimization problem is NP-hard \cite{hochba1997approximation}. In this work, unlike the existing solutions that compute dictionary and representation iteratively (i.e., method of optimal directions, stochastic gradient descent, and least absolute shrinkage and selection operator (LASSO)), we first obtain $\bf D$ by finding the dictionaries of $\bm O$ and then solve \eqref{eq:sparse} as a sparsity constraint linear regression problem by using Nadaraya-Watson kernel regression. To achieve this, a cluster ensemble framework is introduced to combine various clustering algorithms \cite{cluster_ensemble}. Specifically, a density-based spatial clustering of applications with noise method is first utilized to cluster all the data points and then use the cluster centers as the dictionaries, which is tabulated as Algorithm \ref{dbscan_algorithm} \cite{jang2019dbscan++}. Two user-defined hyperparameters, a threshold for the minimum number of neighbors, $\gamma$, and the radius, $\xi$, are utilized to perform a minimum density level estimation. $\bm O_i$ with more than $\gamma$ neighbors within $\xi$ distance are considered to be a centroid. All neighbors within the $\xi$ radius of the centroid are considered to be part of the same group as this centroid. This method is capable of finding clusters with arbitrary shapes and sizes and shows robustness and practicality because it does not require \emph{a priori} specification on the number of clusters. By fixing $\bf D$, the entry of $\bf R$ can be calculated using the following equation:
\begin{equation}
\label{eq:kernel_solution}
r_{ij} = \frac{K({\bm O_i},{\bf D_j})}{\sum_{j'\in{\bf D}} K({\bm O_i},{\bf D_j})}
\end{equation}
where, $K(\cdot,\cdot)$ is a per-defined kernel function. 

After applying the sparse coding, the representation matrix $\bf Z$ can be represented as an undirected similarity graph, $\bf G = (Z,E)$, where $\bf E$ is a set of edge connecting different vertices. The undirected graph $\bf G$ is represented by its adjacency matrix $W = (\mathrm{w}_{i,j})_\{i,j = 1,...,m\}$, where $\mathrm{w}_{i,j}>0$ indicates the similarity between two selected outage samples, $\bm O_i$ and $\bm O_j$. $W_{i,j}$ = 0 indicating that the vertices $\bf Z_i$ and $\bf Z_j$ are not connected. We have adopted the Gaussian kernel function to build the adjacency matrix $W$ as follows:
\begin{equation}
\label{eq:GK}
W_{i,j} = exp\left( \frac{-||\bf Z_i - Z_j||^{2}}{\alpha ^{2}} \right )
\end{equation}
where $\alpha$ is a scaling parameter indicating how fast the weight decreases with the distance between the two vertices $\bf Z_i$ and $\bf Z_j$. To avoid the error caused by manual parameter selection, a localized scaling parameter $\alpha_i$ is calculated for each vertex, which allows self-tuning of the point-to-point distances based on the local distance of the neighbor of $\bf Z_i$ \cite{GJ2007}:
\begin{equation}
\alpha_i = ||\bf Z_i - Z_\beta||
\end{equation}
where $\bf Z_\beta$ is the $\beta$'th neighbor of $\bf Z_i$. Then, the weight equation \eqref{eq:GK} can be reformulated as follows:
\begin{equation}
W_{i,j} = exp\left( \frac{-||\bf Z_i - Z_j||^{2}}{\alpha_i \alpha_j} \right ).
\end{equation}
By defining a couple of vertices and weight matrix $W$, the outage data grouping is converted to a graph partitioning problem. The objective function of the graph partitioning is to maximize both the dissimilarity between the disparate groups and the total similarity within each group:
\begin{equation}
S({\bf G})= \displaystyle  \min_{\bf A_1,A_2,...,A_\theta } \sum_{i=1}^{\theta }\frac{s(\bf A_i,Z\setminus A_i)}{d(\bf A_i)}
\end{equation}
where $\theta$ is the number of vertices, $\bf A_i$ is a group of vertices in $\bf Z$, $\bf (Z\setminus A_i)$ is the nodes of set $\bf Z$ that are not in $\bf A_i$, $ s(\bf A_i,Z\setminus A_i)$ is the sum of the weights of vertices in $\bf A_i$. Then, the Laplacian matrix is formulated based on the weight matrix $W$:
\begin{equation}
    L =  {\bf D}^{-\frac{1}{2}}W{\bf D}^{-\frac{1}{2}}
\end{equation}
where $\bf D$ is a diagonal matrix which the $(i,i)$'th element represents the sum of $W$'s corresponding row. When solving the graph partition problem, according to the \textit{Rayleigh-Ritz Theorem}, the solution is acquired by using $k$ smallest eigenvectors of the Laplacian matrix, which guarantees an approximate value of the \textit{optimal cut} \cite{chung1997,aggarwal2014data}. The value of $k$ can be determined by various clustering evaluation metrics, such as the Silhouette coefficient, Dunn's Index, and Davies-Bouldin validation index (DBI). In this work, to set the optimal $k$, we adopt the DBI, which purposes to minimize the overlap of different groups and maximize the conformance within each group. When we modify the $k$ value for finding the smallest eigenvalues of the Laplacian matrix, the corresponding DBI value is recorded for each $k$. The optimal value of $k$ is determined when the DBI is minimized \cite{dbivalidation}. When the value of $k$ is assigned, a new matrix ${\bf R} \in \mathbb{R}^{n \times k}$ is built based on the $k$ smallest eigenvalues of the $L$ matrix. Based on the properties of the graph Laplacians, the data point $\bf Z_i$ is reconstructed using the $i$’th row of the matrix $\bf{R}$, which enhances the cluster-properties of the data \cite{GJ2007}. After the data reconstruction, any clustering methods can be used to easily obtain the results. In this work, we used $k$-means algorithm to obtain the final solutions from matrix $\bf{R}$.

\begin{algorithm}[t]
\renewcommand\baselinestretch{1}\selectfont
\caption{Dictionary Selection in SDESC}
\begin{algorithmic}[0]\label{dbscan_algorithm}
\STATE \hspace{-3mm}{\bf Initialization}: Initialize $i\leftarrow 1$, $\gamma$, $\xi$
\STATE \hspace{-3mm}{\bf repeat}
\STATE [{\bf S1}]: Select the $i$th row of $\bm O$.
\STATE \hspace{-0mm}{\bf repeat}
\STATE \hspace{3mm}[{\bf S2}]: Pick $\bm O_{i}$ and retrieve all direct density-reachable points in $\bm O$ using $\xi$.\\
\STATE \hspace{3mm}[{\bf S3}]: Based on $\gamma$, if $\bm O_{i}$ is a core point, a cluster is formed; \\ \hspace{11mm}otherwise, assign $\bm O_{i}$ to noise
\STATE [{\bf S4}]: Update $i\leftarrow i+1$.
\STATE \hspace{-3mm}{\bf until} {$i=n$}
\end{algorithmic}
\end{algorithm}

The proposed SDESC algorithm has three unique advantages over the conventional spectral clustering (SC) method in this task: 1) The enhanced cluster-properties of the reconstructed dataset reduce the sensitivity of the clustering process to bad outage records that are unavoidable in real-world applications. 2) The proposed method introduces the dictionary-based weight matrix of the dataset rather than computing the high-dimensional profiles of all available outage data directly. Such a strategy can significantly reduce the complexities of computing the adjacency matrix and graph Laplacian matrix from $O(m^2)$ and $O(m^3)$ to $O(pm)$ and $O(p^{3}+p^{2} m)$, which is beneficial in the big data age. 3) The graph partitioning problem could be settled without making any assumptions on the data distribution. This step enhances the robustness of the clustering method, which leads to better performance for complicated outage data structures \cite{GJ2007}.

\section{Outage Restoration Time Prediction}

Upon the outage data partitioning results, each outage subset is first assigned with an artificial neural network (ANN) to estimate the outage restoration time. As discussed in Section \ref{introduction} and Section \ref{descreption}-B, there may be huge gaps between the amount of data available for different patterns and scales of outage events. However, there are always internal relationships between different outage event patterns. To this end, a transfer-learning-based method is then employed to transfer the learned knowledge from one prediction model to enhance the performance of the rest models. To help the reader understand the proposed model, we first briefly revisit the concept and properties of ANN, then describe the transfer learning strategy in detail.

\subsection{Artificial Neural Network}

An ANN is constructed with outage-related features as input vector $\bm O$, multiple hidden layers, and an output layer with single output whose value can be treated as the sum of the weighted non-linear terms $\bm O'$:
\begin{equation}
   {\bm O'} =  \sum_{j=1}^{h}(T_{ju} * \sigma (y_j))
\end{equation}
where $h$ is the number of hidden layers, $T_{ju}$ is the weight factor between hidden nodes and output node, $\sigma (y_j)$ is a sigmoidal activation function, which is written as:
\begin{equation}
    \sigma (y_j) = \frac{1}{1+exp(-y_j)}.
\end{equation}
Here, $y_j$ is the input to the $j$-th hidden node, which is defined as:
\begin{equation}
    y_j = \sum_{i=1}^{m} T_{ij}(o_i) + b_i
\end{equation}
where $m$ is the total number of input nodes, $T_{ij}$ is the weights factor between input nodes and hidden node, $o_i$ is the input to the $i$-th input node, and $b_i$ is the bias to the $i$-th input node. Then the least square algorithm is applied to train ANN-based regression models to minimized the total estimation error:
\begin{equation}
    \min \varepsilon =\frac{1}{2} \sum_{i=1}^{m} \left ({\bm O}(i) - {\bm O'}(i) \right )^{2}.
\end{equation}
In this work, the Levenberg-Marquardt (LM) backpropagation method is utilized to update the weights and threshold parameters \cite{backpropagation}. The LM method is derived from Newton’s method to minimize sum-of-squares error functions \cite{levenberg}. It can automatically adjust the learning rate in the direction of the gradient using the Hessian matrix. Compared to back propagation methods with a constant learning rate, the LM method significantly boosts the training speed \cite{wilamowski2010improved}. To calibrate the parameters of ANN, the optimal set of hyper-parameter is determined by the grid search method \cite{randomsearch1}.

\subsection{Transfer Learning-added Outage Restoration Time Prediction Model}

When the training targets are multiple related tasks (i.e., restoration time prediction for outages under different scales and factors), conventional machine learning-based methods need to train multiple models from scratch, thus requiring a large and comprehensive dataset. Such a requirement renders their practical implementation costly. In contrast, transfer learning-based models greatly reduce the amount of data required for training by leveraging prior knowledge gained from previous training tasks \cite{transfer}. Therefore, a transfer learning strategy is adopted to discover domain-invariant intrinsic outage features and structures under different but related domains, establishing the re-utilization of data information across domains. 

In this work, \textit{learning tasks} are defined as the training assignments of each outage subset, and the \textit{source task} is defined as the pre-learned prediction model. To determine the optimal initial source task and assess the similarities between the source task and learning tasks, a validation dataset is randomly selected from each subset. The optimal source task is determined by selecting the subset with the highest average similarity with other subsets. Once the source task is settled, a knowledge matrix related to the source task is first obtained during the training process, which is then exploited as the initial knowledge for a new learning task. Such a process can be formulated as:
\begin{equation}
\label{eq:TL}
    {\bm T}_i = \sum \omega {\bm T}^*_i , \hspace{4mm} i = 1,2,...,n
\end{equation}
where ${\bm T}_i$ is the initial knowledge matrix of the $i$-th variable in a learning task; ${\bm T}^*_i$ denotes the learned knowledge matrix of the $i$-th parameter in the source task; $\omega$ represents the similarity between the learning task and the source task; $n$ is the number of data features. 

\begin{figure}[tbp] 
	\centering
	\includegraphics[width=3.4in]{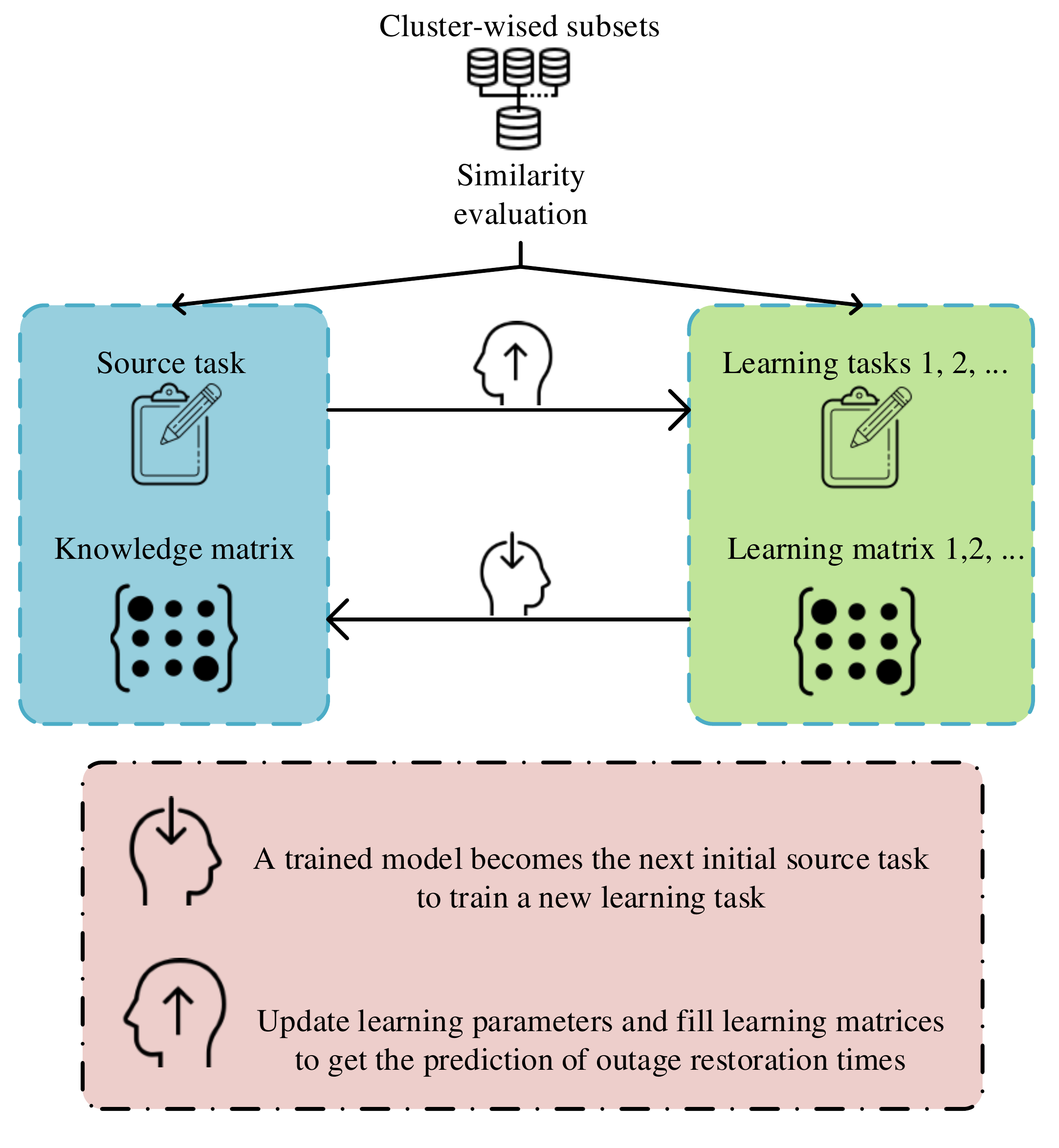}
	\caption{Exemplary of the transfer learning process.}
	\label{fig:TL}
\end{figure}

This transfer learning process is depicted in Fig. \ref{fig:TL}. In detail, after the source task and learning tasks are distinguished by a similarity evaluation of cluster-wised outage subsets, the transfer learning process gathers the outage-related features and the output (i.e., actual restoration time) in the pre-trained model, and stores them as a knowledge matrix. Similarly, each learning task, which is an untrained model in the queue, generates a learning matrix containing outage features, the corresponding actual restoration time, and an empty pending entry that represents the predicted restoration time for each outage. According to \eqref{eq:TL}, by exploiting the similarity $\omega$ between the learning task and the source task, which is now the similarity between the knowledge matrix and learning matrix, the learning parameters can be updated for training a new prediction model. After updating the parameters of the learning matrix, the predicted restoration time can be obtained by directly filling the pending entries in the matrix for each outage. When the first prediction task is completed, the learned model can be utilized in a recursive manner when dealing with a new learning task \cite{transferlearningsurvey}. For example, the training assignments for cluster-wised subsets are signed as task 1, 2 and 3. Task 2 is learned by exploiting task 1 as a source task, then task 2 can be used as the source task for training task 3.

Compared to the conventional machine learning-based method, the proposed transfer learning-added outage restoration time prediction method has two unique advantages: 1) the transfer learning strategy with cluster-wised datasets can greatly reduce the overfitting risk caused by the data scarcity of the specific outage prediction patterns and distributions. This can be confirmed using the numerical results. 2) The transfer learning strategy can introduce better robustness for imperfect data. In actual grids, some outage-related features may be incorrectly collected or recorded, which is difficult to resolve by applying data imputation methods. However, in our method, the similarity parameter $\omega$ is introduced to identify such data. The incorrect data stored in the learning matrix does not have justified similarity with data measurements in the source matrix. Therefore, such data can be easily removed by sorting the similarities for each outage data from the learning task. As the learning task becomes a new source task, incorrect data in the learning task is eventually reduced to a small number by repeating this process.

\subsection{Unseen Outage Classification}

To identify and allocate the corresponding outage pattern and related prediction model to unseen outage, a t-distributed stochastic neighbor embedding (t-SNE) method is utilized. By optimizing the conditional probability between original data and analogue data, the t-SNE can convert high-dimensional outage data into low-dimensional representations. Data with higher similarity in high-dimensional space is closer to each other after being embedded in low-dimensional space, thus leading to better classification performance \cite{tsnereview, 8899670}. Here, the conditional probability between any two outage samples ${\bm O}_i$ and ${\bm O}_j$ in high-dimensional space can be formulated as:
\begin{equation}
    P_{A|B}(\bm O_i|\bm O_j)=\frac{exp \left ( \frac{-||{\bm O}_j - {\bm O}_i||^2}{2{\delta}_j} \right )}{\sum_{\kappa=1,\kappa\neq j}^{m}exp\left (  \frac{-||{\bm O}_j - {\bm O}_\kappa||^2}{2{\delta}_j}\right )}
\end{equation}
\begin{equation}
    P_{A|B}(\bm O_j|\bm O_i)=\frac{exp \left ( \frac{-||{\bm O}_i - {\bm O}_j||^2}{2{\delta}_i} \right )}{\sum_{\kappa=1,\kappa\neq i}^{m}exp\left (  \frac{-||{\bm O}_i - {\bm O}_\kappa||^2}{2{\delta}_i}\right )}
\end{equation}
where $\kappa$ is the number of domain points, ${\delta}_i$ is the vector variance of the Gaussian function centered on the data ${\bm O}_i$. The joint probability of ${\bm O}_i$ and ${\bm O}_j$, $P(A=\bm O_i,B=\bm O_j)$, within a Gaussian space is:
\begin{equation}
P_{A,B}(\bm O_i,\bm O_j) = \frac{P_{A|B}(\bm O_i|\bm O_j)+ P_{A|B}(\bm O_j|\bm O_i)}{2m}.
\end{equation}
In a low-dimensional space, the t distribution is applied with one degree of freedom. The joint distribution of two simulated data $\bm{Y}_i$ and $\bm{Y}_j$ is calculated as follows:
\begin{equation}
P_{A,B}(\bm Y_i,\bm Y_j) = \frac{\left ( 1 + ||\bm{Y}_i - \bm{Y}_j||^2 \right )^{-1}}{\sum_{\kappa=1, \kappa\neq l}^{m} \left ( 1+||\bm{Y}_\kappa - \bm{Y}_l||^2 \right )^{-1}}.
\end{equation}
The Kullback Leibler (KL) divergence is utilized to quantify the similarity between $P_{A,B}(\bm Y_i,\bm Y_j)$ and $P_{A,B}(\bm O_i,\bm O_j)$:
\begin{equation}
{\bm C} = \sum_{i=1}^{m}\sum_{j=1}^{m} P_{A,B}(\bm O_i,\bm O_j) \log_{2} \frac{P_{A,B}(\bm O_i,\bm O_j)}{P_{A,B}(\bm Y_i,\bm Y_j)}.
\end{equation}
The optimal low-dimensional data, ${{\hat{\bm Y}}_1,\hat{\bm{Y}}_2,...,\hat{\bm{Y}}_m}$ is obtained by minimizing the KL divergence using the gradient descent method, which is denoted as:
\begin{equation}
\frac{\delta {\bm C}}{\delta \bm{Y}_i} = 4 \sum_{j=1}^{m} \left ( {P_{ij}} - {q_{ij}}\right ) \left ( \bm{Y}_i - \bm{Y}_j \right )  \left ( 1+||\bm{Y}_i - \bm{Y}_j||^2 \right )^{-1}.
\end{equation}

When an outage occurs, the high-dimensional data is first converted into the two-dimensional map using the t-SNE method. Then, the distances between the edges of each subset with this point are calculated to quantify the similarity between the unseen outage and the learned subsets. Based on the distance values, the most probable class is chosen as the correct underlying subset for the unseen outage.

\section{Numerical Results}\label{result}
This section explores the practical effectiveness of the proposed outage restoration time prediction method. A real-world dataset with 16,000 outages events is utilized in this case study, including six years of data collected by a utility in New York State. After data preprocessing, the whole dataset is randomly divided into three parts for training, testing, and validation by 70\%, 15\%, and 15\% of the total data, respectively. 

\subsection{SDESC Algorithm Performance}

\begin{figure}[tbp] 
	\centering
	\includegraphics[width=3.4in]{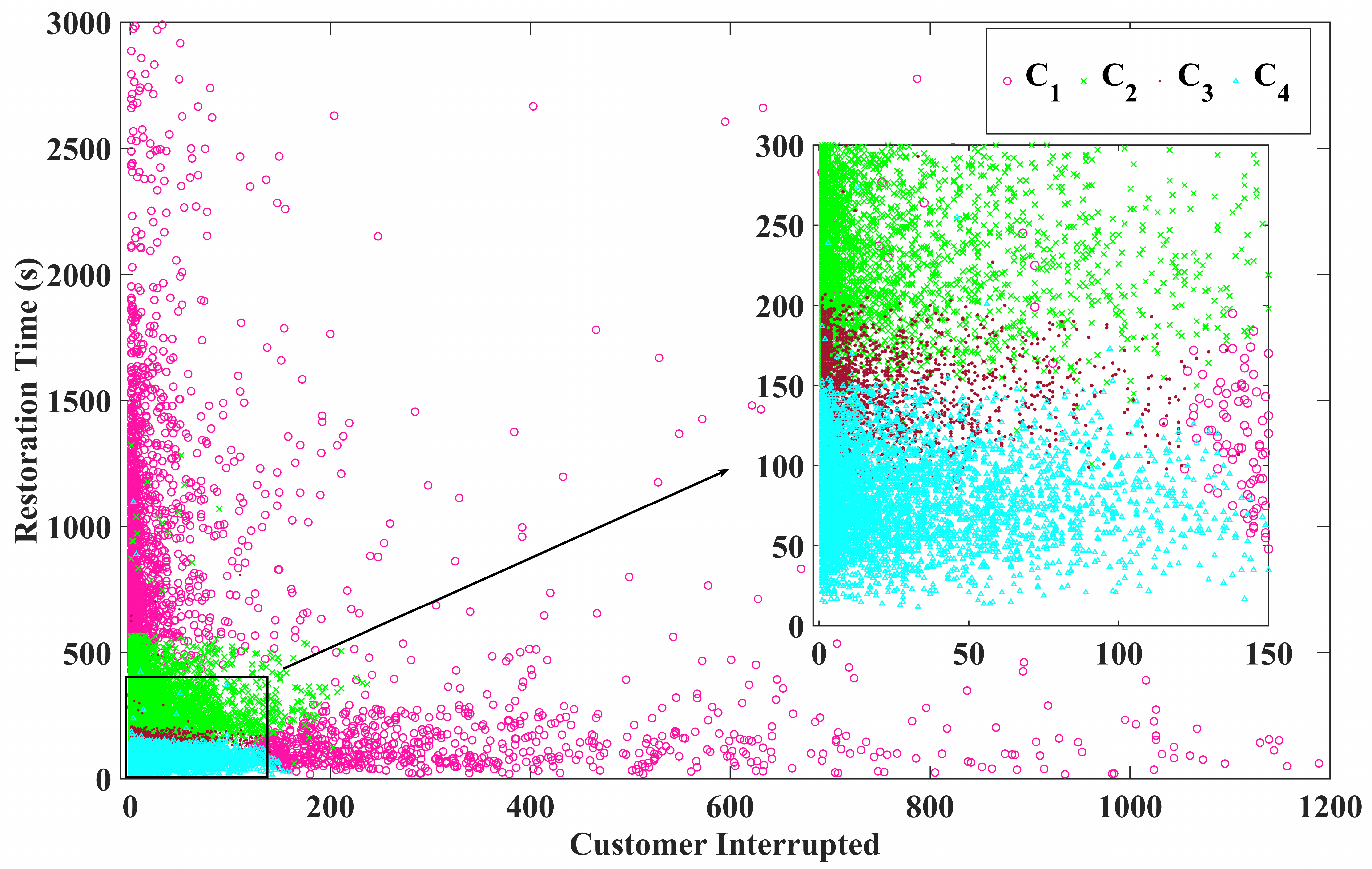}
	\caption{Dictionary-based ensemble spectral clustering of the outage dataset.}
	\label{fig:LSC}
\end{figure}

Fig. \ref{fig:LSC} presents the data distribution of the number of customer interruptions versus outage restoration time. As described in this figure, the number of customers interrupted has a clear impact on restoration time. The rationale behind this is that utilities typically prioritize dealing with large-scale outages; therefore, outage restoration time tends to be shorter when the number of affected customers is higher. In contrast, estimating restoration time for small-scale outages is more challenging, as they are more likely to be affected by a variety of factors. This can also be confirmed using real-world data, as shown in Fig. \ref{fig:LSC}. Therefore, it is necessary to implement the proposed SDESC method to distinguish the outage groups that are constrained by various features.

\begin{figure}[tbp] 
	\centering
	\includegraphics[width=3.4in]{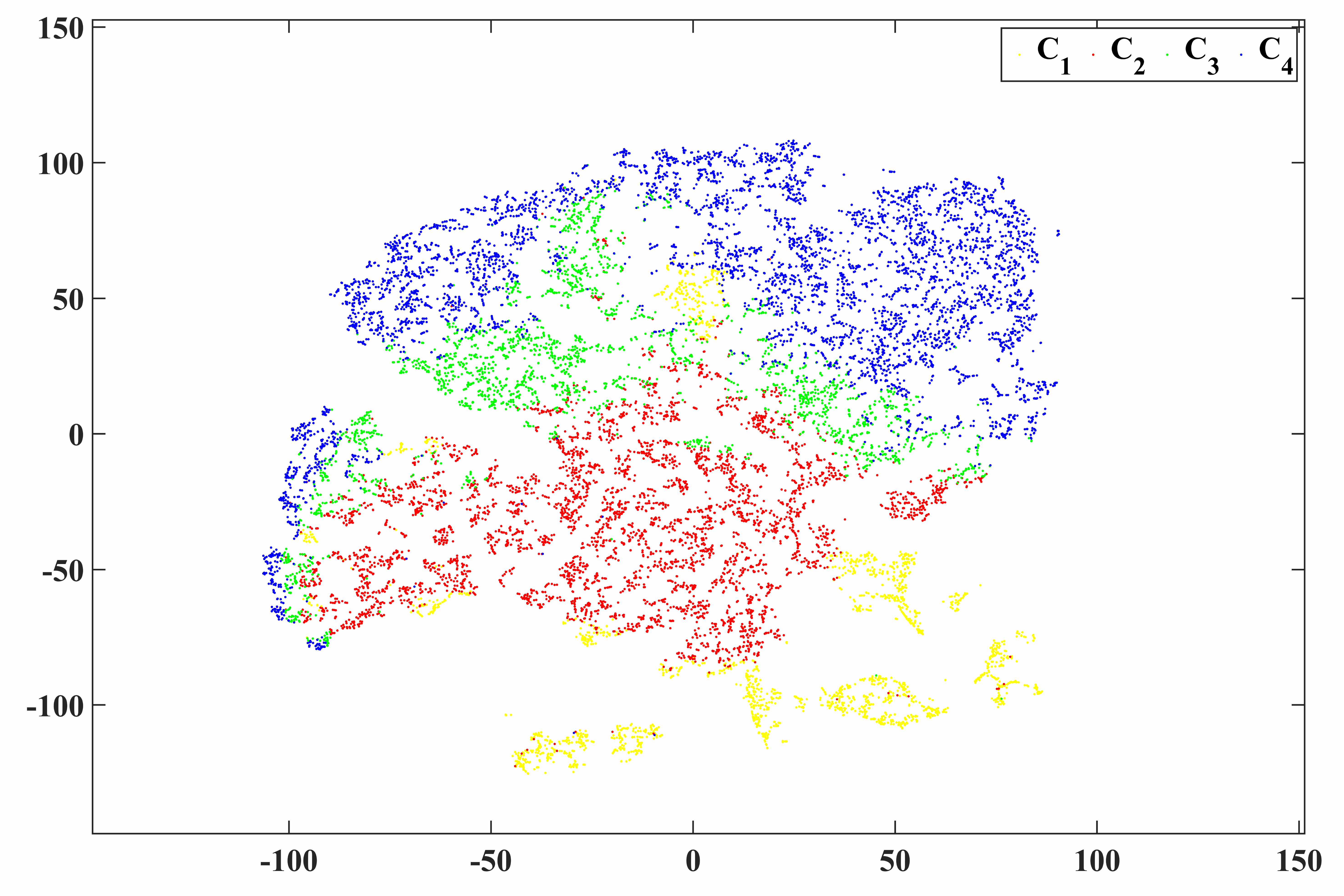}
	\caption{t-SNE plot of clustered data using the proposed SDESC method.}
	\label{fig:lsctsne}
\end{figure}

Basically, the SDESC calibration is a trial and error process using a specific cluster evaluation metric. In this work, the optimal number of subsets, $k$, is assigned as $4$ based on the minimum DBI value. The grouping results with the corresponding $k$ value are marked in Fig. \ref{fig:LSC}. Specifically, each color represents a subset of the outage data, namely $C_1$, $C_2$, $C_3$, and $C_4$. Table \ref{table:cluster} demonstrates the statistics in the grouping results, including the number of data samples in each subset, the average number of customers interrupted (i.e., Avg. CI), and the average restoration time in minutes (i.e., Avg. RT). As shown in table, $\{C_1, C_2, C_3, C_4\}$ consist of $\{2379, 5302, 2884, 5872\}$ outage data samples, respectively. Such results promise the data imbalance problem: subset $C_2$ and $C_4$ have twice as many data samples as $C_1$ and $C_3$, while the average recovery time (i.e., 740.5 minutes) and the number of customers interrupted (i.e., 170) of $C_1$ are significantly higher than for the other subsets. In our view, $C_1$ refers to severe outages with a higher Avg. RT and Avg. CI, but relatively infrequent. $C_2$ and $C_4$ represent intermediate and least serious outages, which are twice as frequent as severe outages. $C_4$ represents a subset of minor outages, which occur frequently but can typically be resolved in a timely manner. For the classification part, the t-SNE result is shown in Fig. \ref{fig:lsctsne}. The shortest distances between the edges of each subset with each unseen outage are calculated to measure the similarity between the unseen outage and the learned patterns. Based on the testing dataset (15\% of the total data), the classification error margin can achieve 5\% when assigning outage patterns to unseen outages.

\begin{table}[btp]
\centering
\caption{CLUSTERING STATISTICS}
\label{table:cluster}
\setlength{\tabcolsep}{6mm}{
\resizebox{\linewidth}{!}{%
\begin{tabular}{cccc} 
\hline\hline
Cluster & Samples & Avg. CI & Avg. RT(min) 
\\[1pt]
\hline 
$C_1$    & 2379    & 170     & 740.5         \\[3pt]
$C_2$    & 5302    & 21      & 288.4         \\[3pt]
$C_3$    & 2884    & 16      & 144.5         \\[3pt]
$C_4$    & 5872    & 22      & 82.2          \\[3pt]
\hline
\end{tabular}
}}
\end{table}

\begin{figure}[tbp] 
	\centering
	\includegraphics[width=3.4in]{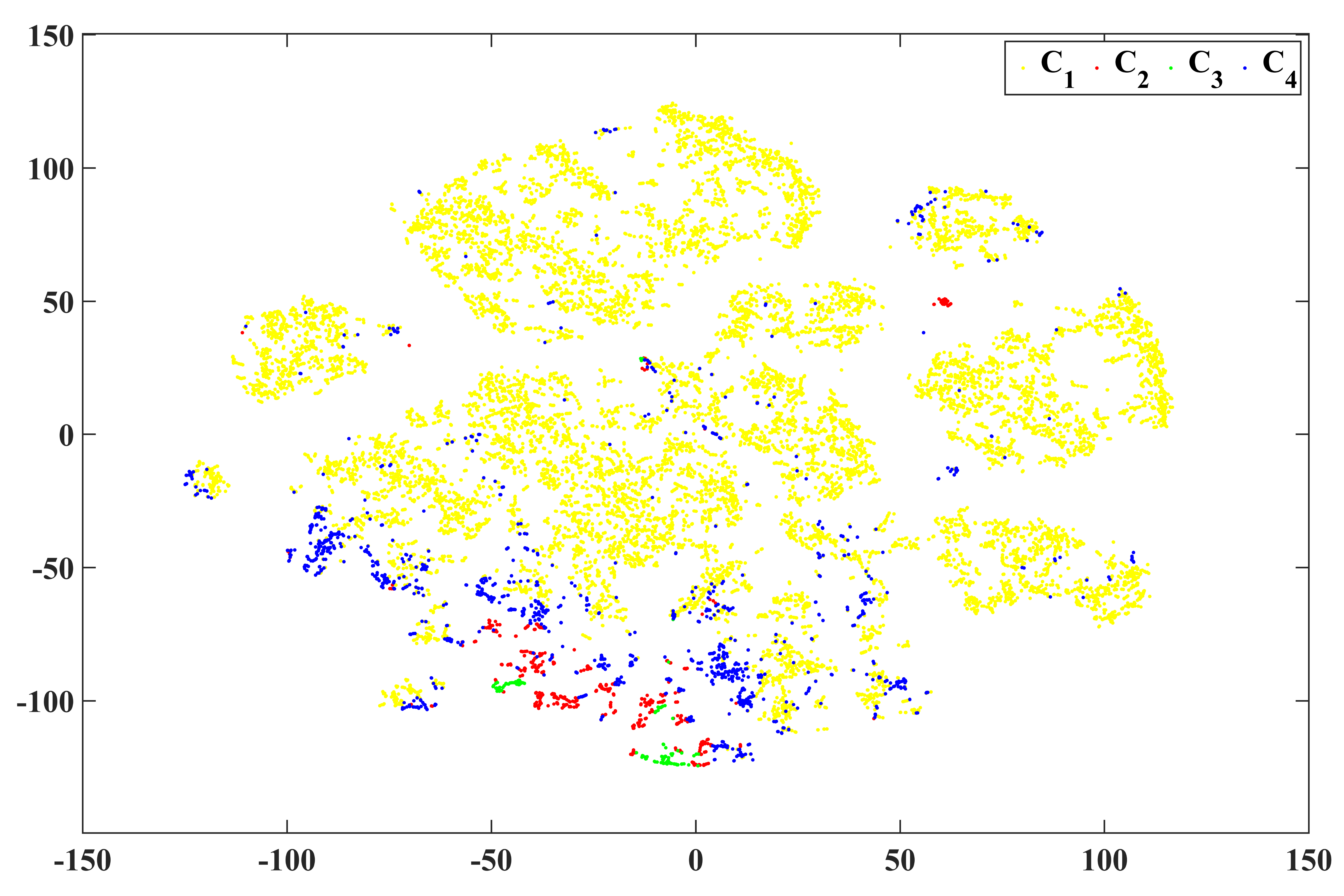}
	\caption{t-SNE plot of clustered data using the advanced K-means method.}
	\label{fig:kmeanstsne}
\end{figure}

We have conducted a numerical comparison with an advanced K-means algorithm \cite{advance_k_means} to show that the proposed SDESC method can offer a dramatic improvement in outage data grouping. Fig. \ref{fig:kmeanstsne} shows the t-SNE plot of the result using the advanced K-means algorithm. By using this state-of-the-art clustering method, over 70\% of the total data has fallen into a single cluster subset, $C_1$. Such a result increases the overfitting risk caused by data insufficient in other subsets. Moreover, it is clear that the data points in different subsets of Fig. \ref{fig:kmeanstsne} are more difficult to be classified than in Fig. \ref{fig:lsctsne}. This indicates that the homogeneity of each subset obtained from the advanced K-means is much lower than that of each subset obtained from the proposed method. Therefore, when an unseen outage occurs, it is highly likely to misclassify, thus resulting in a decrease in restoration time prediction accuracy.

\subsection{Outage Restoration Time Prediction Performance Analysis}

\begin{figure}[tbp]
	\centering
	\includegraphics[width=3.4in]{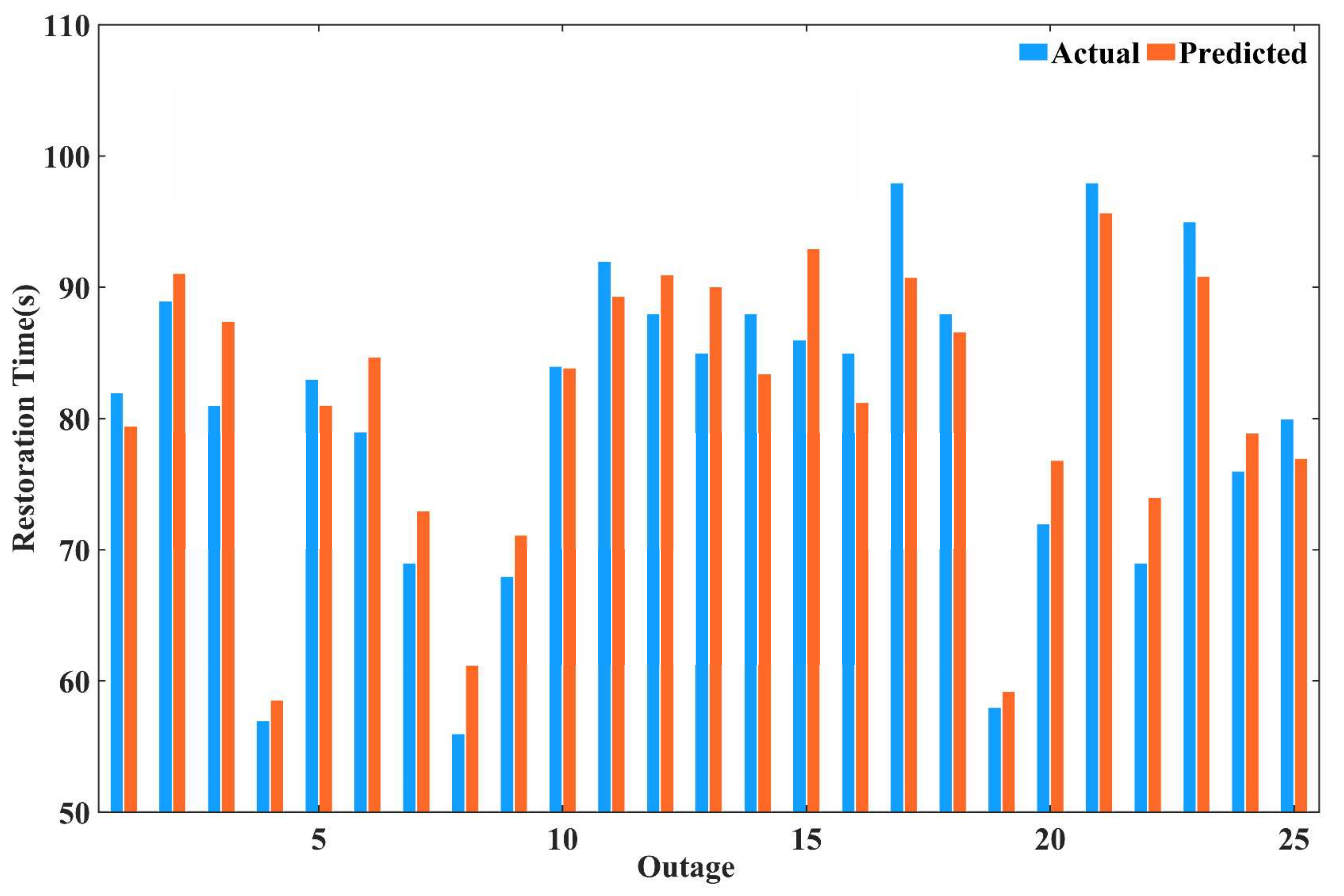}
	\caption{Comparison result between actual and predicted restoration time for the source task ($C_4$).}
	\label{fig:avsp4}
\end{figure}

When the outage dataset is separated using the SDESC method, the ANN with a transfer learning embedded model is utilized to predict the outage restoration time. In this paper, according to the performance of validation sets, we select subset $C_4$ as the source task to initialize the transfer learning strategy. To evaluate the prediction performance of the ANN, the mean absolute percentage error (MAPE) is utilized in this paper, which is formulated as follows:
\begin{equation}
    MAPE = \frac{100}{m}\sum_{i=1}^{m}\left|\frac{A_i-P_i}{A_i} \right|
\end{equation}
where $A_i$ is the actual data value, and $P_i$ is the predicted value, their difference is divided by the actual data value. The absolute value in this division is summed up for every predicted point in time and divided by the number of data measurements $m$. In addition to MAPE, the percentage of predicted restoration time that falls within the reasonable range from the actual restoration time is also calculated to further evaluate the performance of our method. Fig. \ref{fig:avsp4} describes the comparison between the actual and predicted restoration time for 25 randomly selected samples in $C_4$. After predicting the restoration time of the test data, the predicted restoration time range is 22.8 minutes, which is below the 30 minutes threshold. 3\% of the predicted time is more than 60 minutes of the actual repair time, and only one particular outage showed that the predicted time is more than 90 minutes of the actual repair time. 

\begin{figure} [tbp]
     \centering
     \begin{subfigure}[h]{0.45\textwidth}
         \centering
         \includegraphics[width=\textwidth]{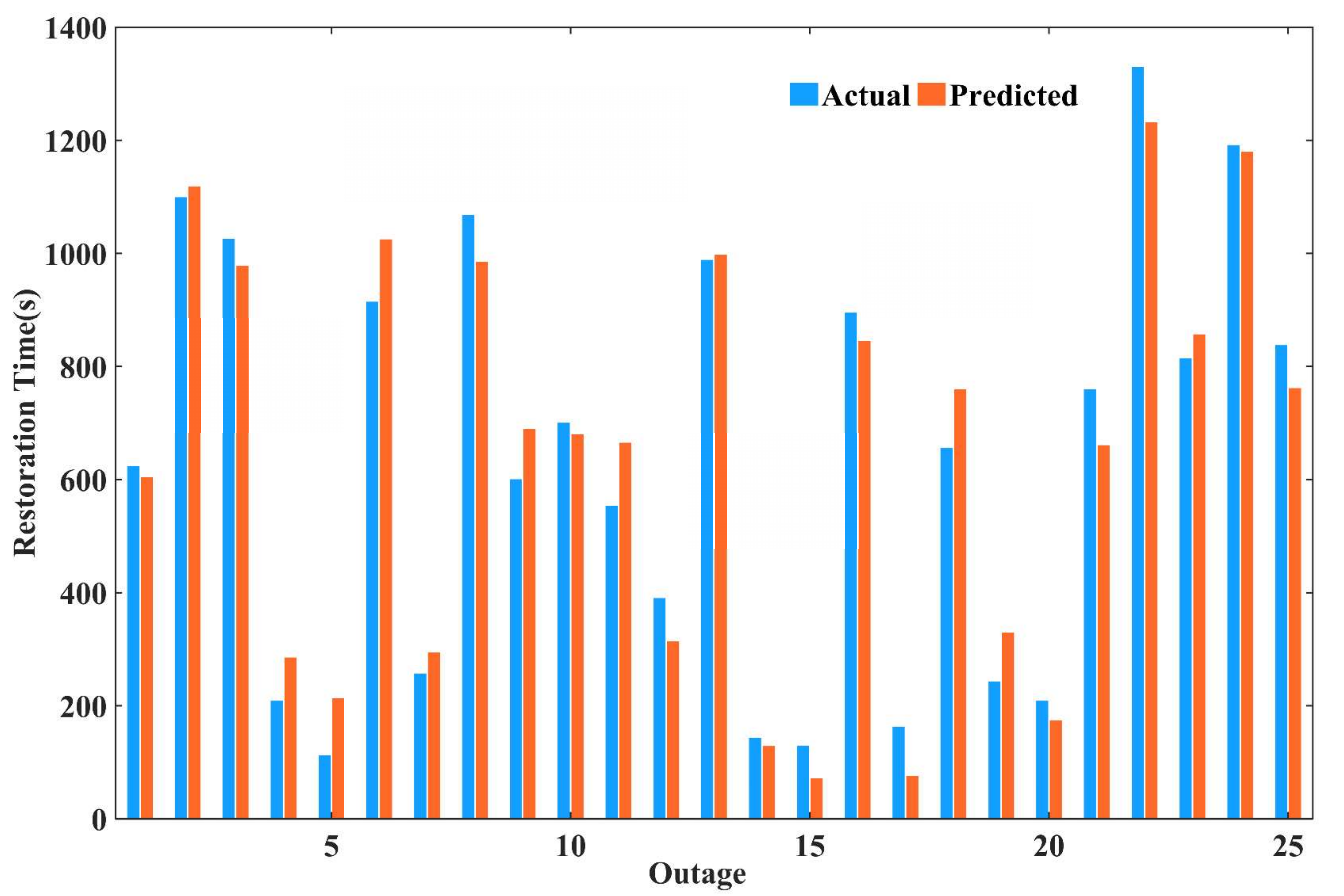}
         \caption{Cluster 1 ($C_1$)}
         \label{fig:avsp1}
     \end{subfigure}
     \hfill
     \begin{subfigure}[h]{0.45\textwidth}
         \centering
         \includegraphics[width=\textwidth]{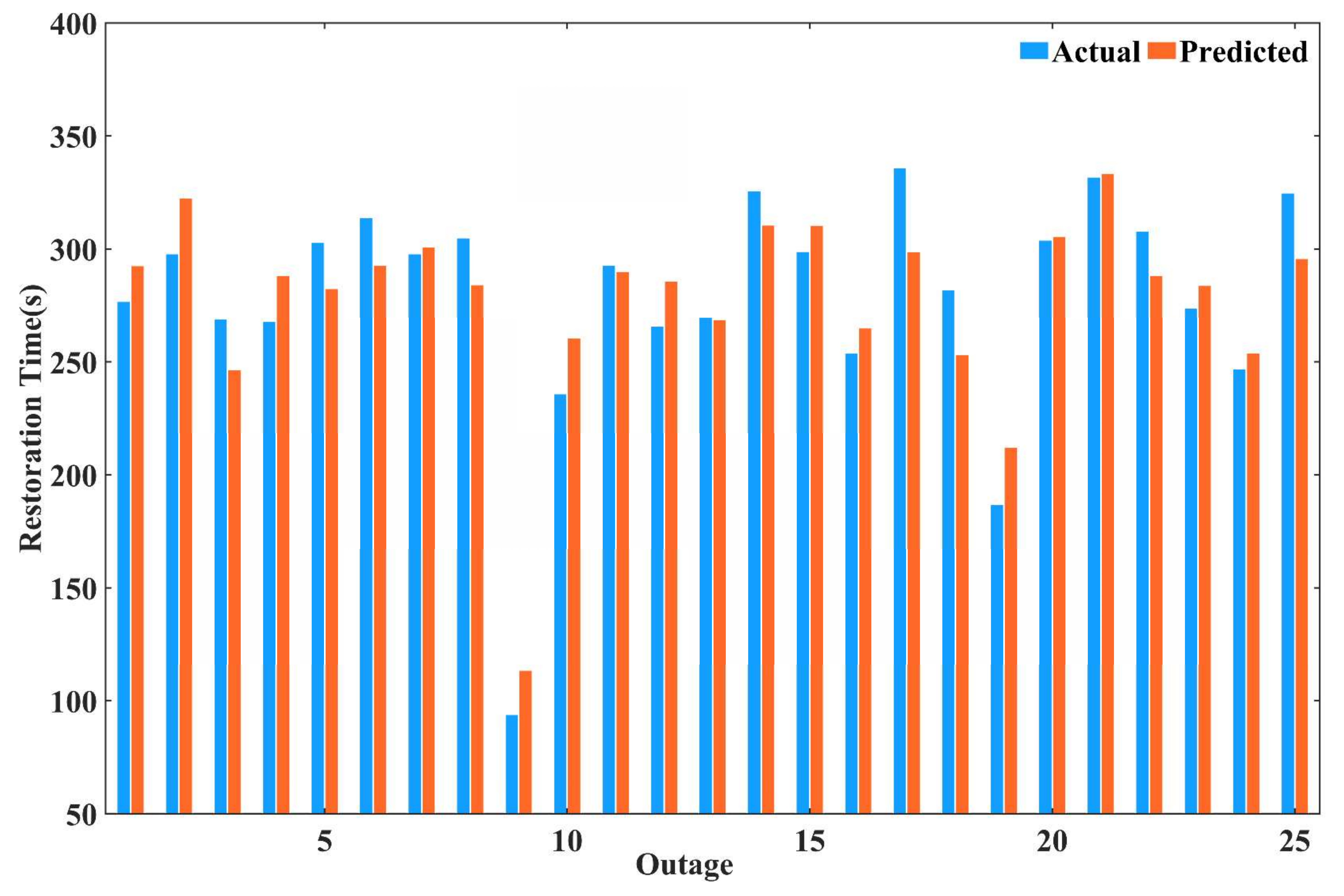}
         \caption{Cluster 2 ($C_2$)}
         \label{fig:avsp2}
     \end{subfigure}
     \hfill
     \begin{subfigure}[h]{0.45\textwidth}
         \centering
         \includegraphics[width=\textwidth]{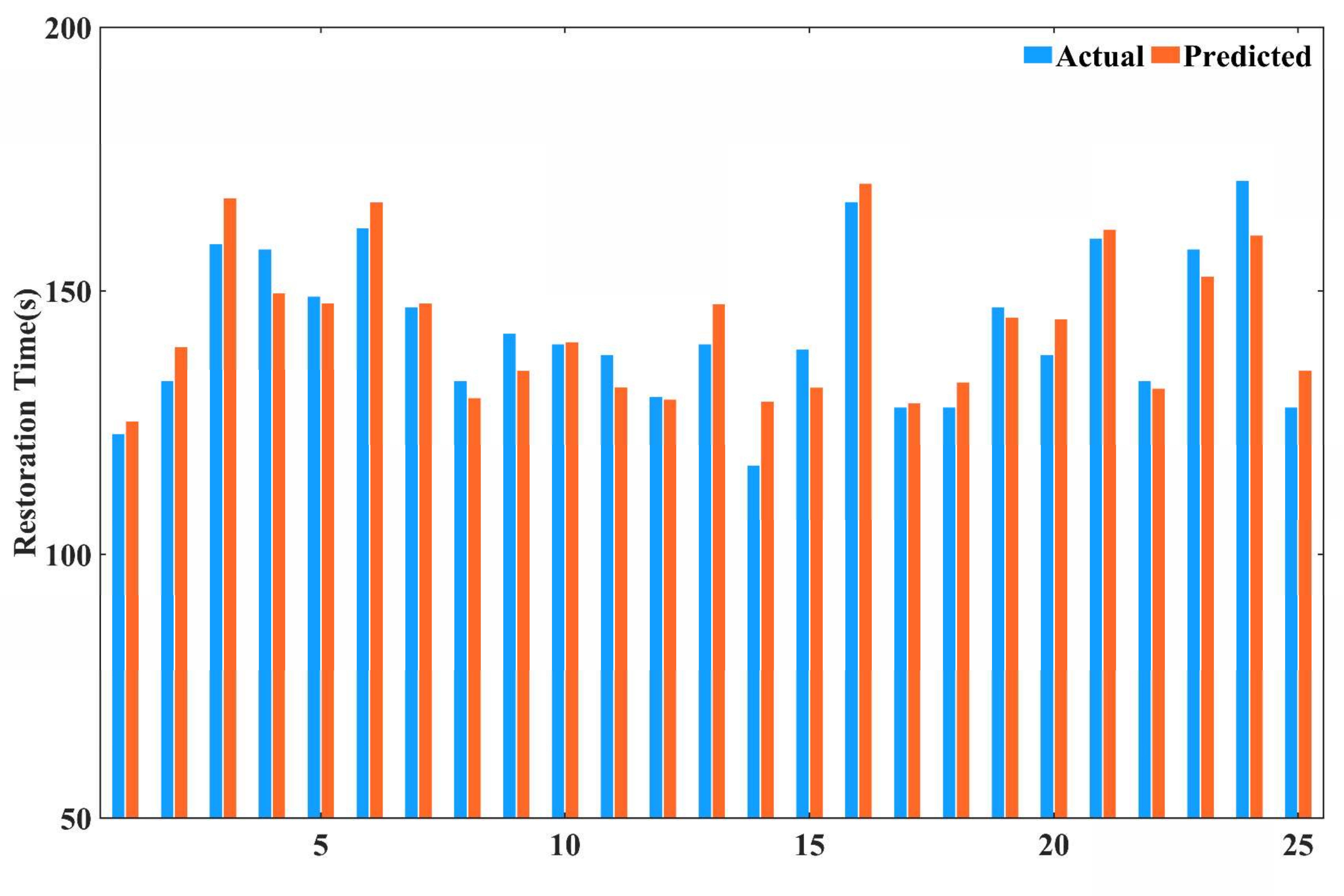}
         \caption{Cluster 3 ($C_3$)}
         \label{fig:avsp3}
     \end{subfigure}
        \caption{Comparison result between actual and predicted restoration time for learning tasks.}
        \label{fig:compareactualpredict}
\end{figure}

The proposed method is utilized to train the prediction models for $C_1$, $C_2$ and $C_3$ using the source task. The MAPE for $C_1$, $C_2$, and $C_3$ is 23\%, 24\%, and 11.7\%, respectively. The predicted restoration time range for all three subsets is 48.8 minutes, while subset $C_3$ had an outstanding prediction range of 21.9 minutes. Fig. \ref{fig:compareactualpredict} demonstrates the comparison between the actual and predicted restoration time for 25 randomly selected outages in $C_1$, $C_2$, and $C_3$. Note that the variance in restoration time for each subset is 1300 seconds, 350 seconds, and 170 seconds. Despite the high variance of $C_1$, the prediction model still performs decently. The prediction accuracy of $C_1$ is slightly lower than that of $C_2$ and $C_3$. This result is reasonable because higher data variance usually leads to a higher risk of overfitting and reduces the accuracy of the prediction model.

\subsection{Method Comparison}

\subsubsection{Proposed Model vs. Global Model}

\begin{figure}[tbp]
	\centering
	\includegraphics[width=3.4in]{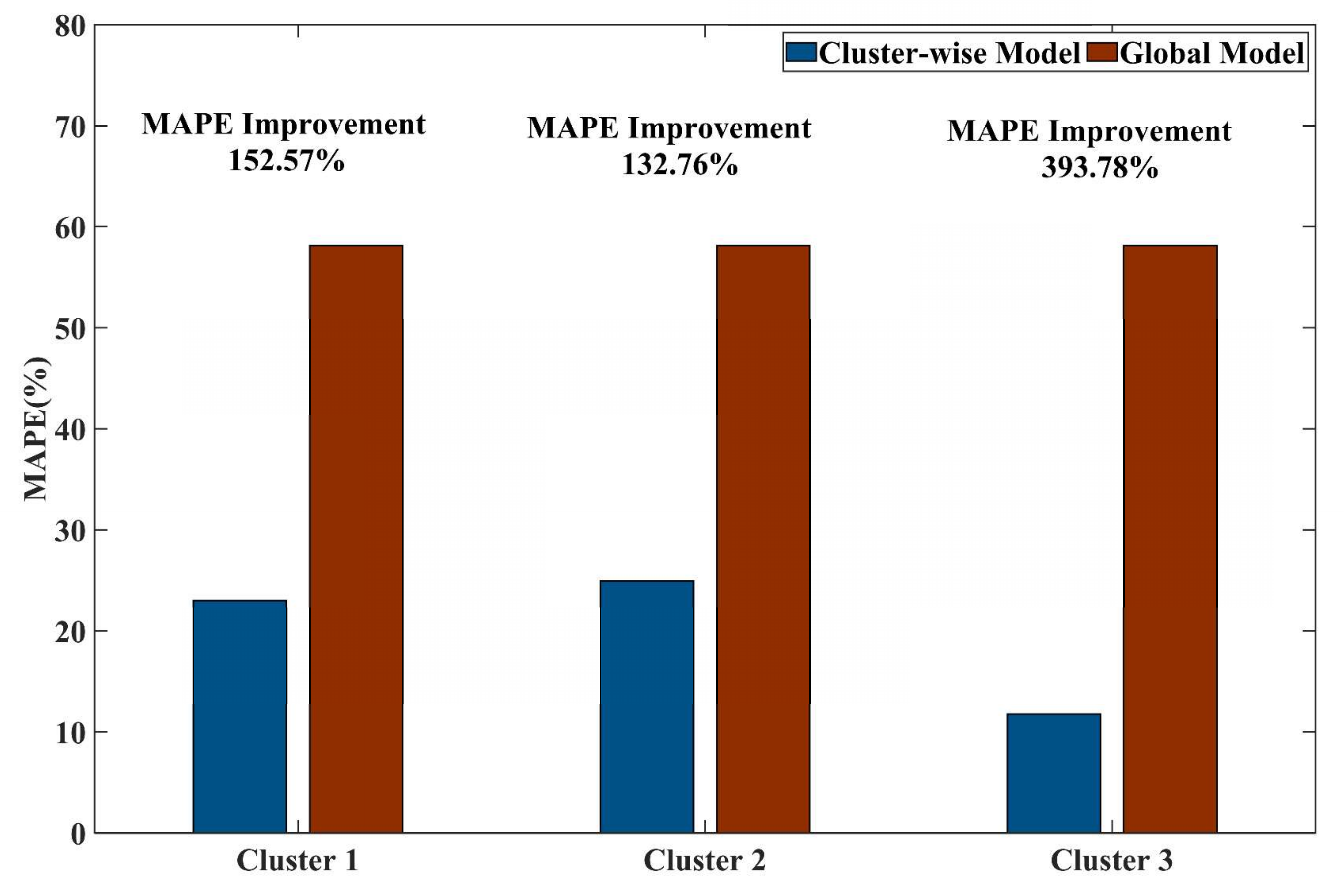}
	\caption{Restoration time comparison between cluster-wised model and global model.}
	\label{fig:cvsg}
\end{figure}

We have conducted a comprehensive comparison between the proposed cluster-wised model with the previous restoration time prediction model\cite{predictiondeeplearning}. Note that the previous model follows a global training fashion and is developed using all outage records without clustering. Where possible we attempted to tune parameters for each algorithm to give a fair comparison. The MAPE improvement compared to the global model for each of the subsets are 152.57\%, 132.76\%, and 393.78\%, respectively, as shown in Fig. \ref{fig:cvsg}. This result indicates that our SDESC method can substantially improve the prediction performance by reducing the uncertainty of the real-world outage data compared to the global model.

\begin{figure}[tbp]
	\centering
	\includegraphics[width=3.4in]{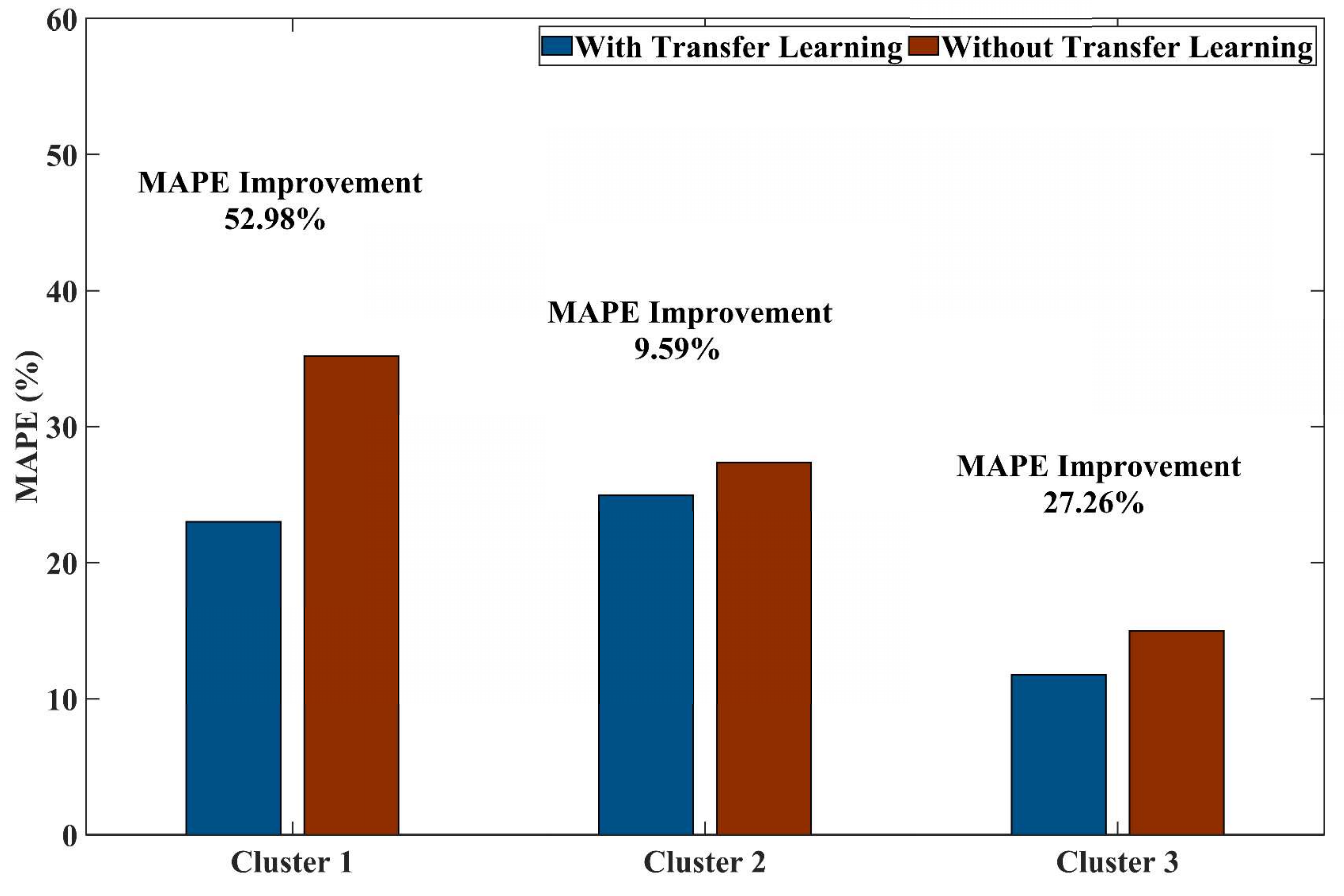}
	\caption{Prediction of restoration time with and w/o transfer learning approach.}
	\label{fig:tlvsnotl}
\end{figure}

\subsubsection{Proposed Model vs. Conventional cluster-wise Learning-based Model}

Another numerical comparison has been conducted between the proposed method with the conventional cluster-wise-based method. This conventional method trained independent neural networks for each cluster-wised subset. Such a comparison can further demonstrate that our method can achieve good prediction performance. Both methods are evaluated based on the same neural network configurations to ensure a fair comparison. As is demonstrated in Fig. \ref{fig:tlvsnotl}, for the three different outage subsets $C_1$, $C_2$, and $C_3$, compared to the conventional cluster-wise-based method, our transfer-learning-added model has 52.98\%, 9.59\%, and 27.26\% MAPE improvement, respectively. The results show that the transfer learning strategy can meet the challenges posed by real-world unbalanced outage datasets and greatly improve the accuracy of repair time prediction.

\section{Conclusion}\label{conclusion}

This paper presents a novel data-driven approach to accurately predict outage restoration time using transfer learning with cluster ensembles. In this paper, six years of real-world outage dataset from our utility partner is investigated for model development and validation. The proposed SDESC approach utilizes the sparse coding technique and cluster ensemble mechanism to first decompose the large-scale datasets, which has good computational efficiency and scalability. Based on the learned outage patterns, the developed transfer-learning-added model can not only accurately predict the outage restoration time in each subset, but also addresses two fundamental challenges: 1) neglect the uncertainty caused by the heterogeneity of outage events with different scales and factors; 2) data imbalance problem in different data subsets. Based on the available real-world utility data, the results show that the proposed method has improved performance compared to existing methods and overcome large-scale data challenges.


%




\ifCLASSOPTIONcaptionsoff
  \newpage
\fi



\bibliographystyle{IEEEtran}
\bibliography{IEEEabrv,./bibtex/bib/IEEEexample}

\begin{thebibliography}{10}
\providecommand{\url}[1]{#1}
\csname url@samestyle\endcsname
\providecommand{\newblock}{\relax}
\providecommand{\bibinfo}[2]{#2}
\providecommand{\BIBentrySTDinterwordspacing}{\spaceskip=0pt\relax}
\providecommand{\BIBentryALTinterwordstretchfactor}{4}
\providecommand{\BIBentryALTinterwordspacing}{\spaceskip=\fontdimen2\font plus
\BIBentryALTinterwordstretchfactor\fontdimen3\font minus
  \fontdimen4\font\relax}
\providecommand{\BIBforeignlanguage}[2]{{%
\expandafter\ifx\csname l@#1\endcsname\relax
\typeout{** WARNING: IEEEtran.bst: No hyphenation pattern has been}%
\typeout{** loaded for the language `#1'. Using the pattern for}%
\typeout{** the default language instead.}%
\else
\language=\csname l@#1\endcsname
\fi
#2}}
\providecommand{\BIBdecl}{\relax}
\BIBdecl

\bibitem{jaech2018real}
A.~Jaech, B.~Zhang, M.~Ostendorf, and D.~S. Kirschen, ``Real-time prediction of
  the duration of distribution system outages,'' \emph{IEEE Trans. Power
  Syst.}, vol.~34, no.~1, pp. 773--781, July. 2018.

\bibitem{modelingweatherrelated}
Y.~Zhou, A.~Pahwa, and S.~S. Yang, ``Modeling weather-related failures of
  overhead distribution lines,'' \emph{IEEE Trans. Power Syst.}, vol.~21,
  no.~4, pp. 1683--1690, October. 2006.

\bibitem{outageevents}
F.~Yang, P.~Watson, M.~Koukoula, and E.~N. Anagnostou, ``Enhancing
  weather-related power outage prediction by event severity classification,''
  \emph{IEEE Access}, vol.~8, pp. 60\,029--60\,042, March. 2020.

\bibitem{adaboost}
P.~Kankanala, S.~Das, and A.~Pahwa, ``Adaboost$^{+}$: An ensemble learning
  approach for estimating weather-related outages in distribution systems,''
  \emph{IEEE Trans. Power Syst.}, vol.~29, no.~1, pp. 359--367, January. 2014.

\bibitem{outagecause}
M.~S. Bashkari, A.~Sami, and M.~Rastegar, ``Outage cause detection in power
  distribution systems based on data mining,'' \emph{IEEE Trans. Ind.
  Informat.}, vol.~17, no.~1, pp. 640--649, 2021.

\bibitem{timeofoutage}
M.-Y. Chow, L.~Taylor, and M.-S. Chow, ``Time of outage restoration analysis in
  distribution systems,'' \emph{IEEE Trans. Power Del.}, vol.~11, no.~3, pp.
  1652--1658, 1996.

\bibitem{liu2007statistical}
H.~Liu, R.~A. Davidson, and T.~V. Apanasovich, ``Statistical forecasting of
  electric power restoration times in hurricanes and ice storms,'' \emph{IEEE
  Trans. Power Syst.}, vol.~22, no.~4, pp. 2270--2279, 2007.

\bibitem{predictiondeeplearning}
A.~Arif and Z.~Wang, ``Distribution network outage data analysis and repair
  time prediction using deep learning,'' \emph{2018 IEEE International
  Conference on Probabilistic Methods Applied to Power Systems (PMAPS)}, pp.
  1--6, June, 2018.

\bibitem{yue2017bayesian}
M.~Yue, T.~Toto, M.~P. Jensen, S.~E. Giangrande, and R.~Lofaro, ``A bayesian
  approach-based outage prediction in electric utility systems using radar
  measurement data,'' \emph{IEEE Trans. Smart Grid}, vol.~9, no.~6, pp.
  6149--6159, November, 2018.

\bibitem{domijan2005effects}
A.~Domijan~Jr, R.~Matavalam, A.~Montenegro, W.~Wilcox, Y.~Joo, L.~Delforn,
  J.~Diaz, L.~Davis, and J.~Agostini, ``Effects of norman weather conditions on
  interuptions in distribution systems,'' \emph{International Journal of Power
  \& Energy Systems}, vol.~25, no.~1, pp. 54--61, 2005.

\bibitem{kankanala2012estimation}
P.~Kankanala, A.~Pahwa, and S.~Das, ``Estimation of overhead distribution
  system outages caused by wind and lightning using an artificial neural
  network,'' in \emph{International Conference on Power System Operation \&
  Planning}, vol. 545, 2012.

\bibitem{owerko2018predicting}
D.~Owerko, F.~Gama, and A.~Ribeiro, ``Predicting power outages using graph
  neural networks,'' \emph{2018 IEEE Global Conference on Signal and
  Information Processing (GlobalSIP)}, pp. 743--747, 2018.

\bibitem{noaasevere}
{National Oceanic and Atmospheric Administration}, ``Severe weather,'' 2021.
  [Online]. Available: {\url{https://www.ncei.noaa.gov/products/severe-weather,
  }}.

\bibitem{noaa}
------, ``Climate data online,'' 2021. [Online]. Available:
  {\url{https://www.ncdc.noaa.gov/cdo-web/}}.

\bibitem{faultcausewithimbalanced}
L.~Xu, M.-Y. Chow, and L.~S. Taylor, ``Power distribution fault cause
  identification with imbalanced data using the data mining-based fuzzy
  classification $e$-algorithm,'' \emph{IEEE Trans. Power Syst.}, vol.~22,
  no.~1, pp. 164--171, 2007.

\bibitem{LSC}
X.~Chen and D.~Cai, ``Large scale spectral clustering with landmark-based
  representation,'' in \emph{Twenty-fifth AAAI Conference on Artificial
  Intelligence}, August. 2011.

\bibitem{hochba1997approximation}
D.~S. Hochba, ``Approximation algorithms for np-hard problems,'' \emph{ACM
  Sigact News}, vol.~28, no.~2, pp. 40--52, 1997.

\bibitem{cluster_ensemble}
J.~Azimi and X.~Fern, ``Adaptive cluster ensemble selection,'' in
  \emph{Twenty-First International Joint Conference on Artificial
  Intelligence}, 2009.

\bibitem{jang2019dbscan++}
J.~Jang and H.~Jiang, ``Dbscan++: Towards fast and scalable density
  clustering,'' \emph{International Conference on Machine Learning}, pp.
  3019--3029, 2019.

\bibitem{GJ2007}
U.~Luxburg, ``A tutorial on spectral clustering,'' \emph{Statistics and
  Computing}, vol.~17, no.~4, pp. 395--416, Mar. 2007.

\bibitem{chung1997}
F.~R. Chung and F.~C. Graham, ``Spectral graph theory,'' \emph{American
  Mathematical Soc.}, no.~92, 1997.

\bibitem{aggarwal2014data}
C.~C. Aggarwal and C.~K. Reddy, ``Data clustering,'' \emph{Algorithms and
  applications. Chapman\&Hall/CRC Data mining and Knowledge Discovery series,
  Londra}, 2014.

\bibitem{dbivalidation}
J.~Bezdek and N.~Pal, ``Some new indexes of cluster validity,'' \emph{IEEE
  Trans. Syst., Man, Cybern. B. Cybern}, vol.~28, no.~3, pp. 301--315, 1998.

\bibitem{backpropagation}
S.~Sapna, A.~Tamilarasi, M.~P. Kumar \emph{et~al.}, ``Backpropagation learning
  algorithm based on levenberg marquardt algorithm,'' \emph{Comp Sci Inform
  Technol (CS and IT)}, vol.~2, pp. 393--398, 2012.

\bibitem{levenberg}
C.~Lv, Y.~Xing, J.~Zhang, X.~Na, Y.~Li, T.~Liu, D.~Cao, and F.-Y. Wang,
  ``Levenberg--marquardt backpropagation training of multilayer neural networks
  for state estimation of a safety-critical cyber-physical system,'' \emph{IEEE
  Trans. Ind. Informat.}, vol.~14, no.~8, pp. 3436--3446, 2017.

\bibitem{wilamowski2010improved}
B.~M. Wilamowski and H.~Yu, ``Improved computation for levenberg--marquardt
  training,'' \emph{IEEE Trans. Neural Netw.}, vol.~21, no.~6, pp. 930--937,
  2010.

\bibitem{randomsearch1}
J.~Bergstra and Y.~Bengio, ``Random search for hyper-parameter optimization,''
  \emph{Journal of Machine Learning Research}, vol.~13, pp. 281--305, Feb.
  2012.

\bibitem{transfer}
L.~Torrey and J.~Shavlik, ``Transfer learning,'' \emph{Handbook of Research on
  Machine Learning Applications and Trends: Algorithms, Methods, and
  Techniques}, pp. 242--264, 2010.

\bibitem{transferlearningsurvey}
F.~Zhuang, Z.~Qi, K.~Duan, D.~Xi, Y.~Zhu, H.~Zhu, H.~Xiong, and Q.~He, ``A
  comprehensive survey on transfer learning,'' \emph{Proceedings of the IEEE},
  vol. 109, no.~1, pp. 43--76, 2021.

\bibitem{tsnereview}
A.~C. Belkina, C.~O. Ciccolella, R.~Anno, R.~Halpert, J.~Spidlen, and J.~E.
  Snyder-Cappione, ``Automated optimized parameters for t-distributed
  stochastic neighbor embedding improve visualization and analysis of large
  datasets,'' \emph{Nature Communications}, vol.~10, no.~1, pp. 1--12, 2019.

\bibitem{8899670}
X.~Fang, S.~Zhang, X.~Su, B.~Zhao, W.~Xiao, Y.~Yin, and F.~Wang, ``Blast
  furnace condition data clustering based on combination of t-distributed
  stochastic neighbor embedding and spectral clustering,'' \emph{2019 IEEE 15th
  International Conference on Control and Automation (ICCA)}, pp. 1608--1613,
  2019.

\bibitem{advance_k_means}
R.~Yadav and A.~Sharma, ``Advanced methods to improve performance of k-means
  algorithm: A review,'' \emph{Global Journal of Computer Science and
  Technology}, vol.~12, no.~9, pp. 47--52, 2012.

\end{thebibliography}

\end{document}